\pdfoutput=1
\documentclass[11pt]{article}
\usepackage[letterpaper,margin=1in]{geometry}
\usepackage[T1]{fontenc}
\usepackage{mathptmx}
\usepackage{courier}
\usepackage{amsmath,amssymb,amsthm}
\allowdisplaybreaks
\usepackage{booktabs}
\usepackage{graphicx}
\usepackage{microtype}
\usepackage[font=small,labelfont=bf]{caption}
\usepackage[authoryear,round]{natbib}
\usepackage[hidelinks]{hyperref}
\hypersetup{pdftitle={Expected Shortfall under the Asymmetric Laplace Working Likelihood in Bayesian Quantile Regression},pdfauthor={Darshan Giri and Songfeng Zheng}}

\newcommand{\VaR}{\mathrm{VaR}}
\newcommand{\ES}{\mathrm{ES}}
\newcommand{\CVaR}{\mathrm{CVaR}}
\newcommand{\AL}{\mathrm{AL}}
\newcommand{\ind}{\mathbf{1}}
\newcommand{\E}{\mathbb{E}}
\newcommand{\Rd}{\eta}
\newcommand{\repo}{\url{https://github.com/dgiri4132/Bayesian-Research}}

\newtheorem{proposition}{Proposition}
\newtheorem{lemma}{Lemma}
\newtheorem{corollary}{Corollary}

\title{\bfseries Expected Shortfall under the Asymmetric Laplace\\
Working Likelihood in Bayesian Quantile Regression}

\author{Darshan Giri\quad and\quad Songfeng Zheng\\
\small Department of Mathematics, Missouri State University}
\date{\small November 2026}

\begin{document}
\maketitle

\begin{abstract}
\noindent
Bayesian quantile regression usually uses an asymmetric Laplace (AL) working
likelihood, and it can be used to estimate Value-at-Risk (VaR), the $\tau$-quantile of the
returns for a small $\tau$. Each posterior draw of the AL's location $\omega$ and scale $\sigma$
also gives an expected shortfall (ES), the mean return at or below VaR, which under the
AL is $\omega-\sigma/(1-\tau)$, so the same fit gives a credible interval for ES. For a
continuous distribution ES equals the conditional Value-at-Risk (CVaR), the name we use
for its true and historical values. We show that this interval is centred in the wrong
place, although the posterior for VaR is centred close to the historical VaR. For daily
SPY returns ($n=2512$, $\tau=0.05$), the posterior mean VaR is $-0.017353$ against a
historical VaR of $-0.017444$, $0.29$ posterior standard deviations away, while the
posterior mean ES is $-0.018868$ against a historical CVaR of $-0.028094$, $29.7$
posterior standard deviations away. We trace this to
two exact properties of the AL: its lower tail is exponential, so the expected loss
beyond VaR is the constant $\sigma/(1-\tau)$, and the fitted $\sigma$ is the minimised
mean check loss, $35\%$ of which comes from the returns below VaR, against $5\%$ under the
AL. In a simulation study,
95\% credible intervals for ES have coverage $95.9\%$ when the
AL is the true model and $0.0\%$ under Gaussian, Student-$t$ and bootstrap designs. We
also define a diagnostic $\Rd$, the ratio of the VaR--CVaR gap implied by a fitted AL to
the true gap, which needs only the mean, VaR and CVaR of a distribution. It equals $1$
when the returns follow the AL with the
same $\tau$, is $0.1422$ for SPY and $0.2598$ for Gaussian returns, and is below $1$ for
all $492$ series in an S\&P~500 cross-section. In a quantile regression with volatility
covariates, the realised mean loss beyond VaR is still about six times what the AL
implies.
\end{abstract}

\section{Introduction}

Value-at-Risk (VaR) and expected shortfall (ES) are the two standard measures of
downside risk. At a small level $\tau$, VaR is the $\tau$-quantile of the returns and ES is
the mean return at or below it, and for a continuous return distribution ES is the same as
what \citet{ru2000} call the conditional Value-at-Risk (CVaR). The Basel Committee's revised
market-risk framework made ``a shift from Value-at-Risk (VaR) to an Expected Shortfall
(ES) measure of risk under stress'' \citep{basel2016}. Quantile regression
\citep{kb1978} estimates VaR without assuming a distribution for the returns.
\citet{ym2001} made it Bayesian by noting that the check loss $\rho_\tau$ is, up to
constants, the negative log density of an asymmetric Laplace (AL) distribution, so an AL
likelihood for the residuals turns quantile regression into a posterior computation.
\citet{kk2011} then wrote the AL as a location--scale mixture, which lets the posterior be
sampled with a simple Gibbs sampler. The AL is used as a working likelihood, not as a
model for returns, because its location parameter is the $\tau$-quantile, and
\citet{srg2013} show that, under mild conditions, the posterior for the regression
parameters concentrates on the true quantile even when the AL is misspecified.

Even so, the AL is a fully specified distribution, and its expected shortfall has a
closed form in the two parameters the sampler produces (Section~\ref{sec:es}):
\begin{equation}
\ES_\tau \;=\; \omega - \frac{\sigma}{1-\tau}.
\label{eq:es}
\end{equation}
So each posterior draw $(\omega^{(g)},\sigma^{(g)})$ also gives a draw of ES, and a
credible interval for ES can be read off a VaR fit without any further modelling. We ask
how this ES behaves when the AL is not the distribution of the returns, which is the
situation in which the AL is used.

We find that the same misspecification affects the two estimates very differently. For ten years of daily SPY returns, the posterior for VaR is centred close
to the historical VaR, while the posterior for ES is $29.7$ posterior standard deviations
away from the historical CVaR (Section~\ref{sec:spy}). The discrepancy has two
sources, and both are exact properties of the AL. First, below its location the AL
density is exponential, so the expected loss beyond VaR is the constant
$\sigma/(1-\tau)$ at every depth in the tail, and the AL's ES is its VaR minus a
constant (Proposition~\ref{prop:flat}). Second, the fitted $\sigma$ is the minimised mean
check loss, which splits as $\hat\sigma=\tau(1-\tau)(e_L+e_U)$ (Lemma~\ref{lem:sigma}). Here $e_L$ is the mean distance below VaR of the
returns at or below it, which is the quantity ES depends on, and $e_U$ is the mean
distance above VaR of the returns above it. For SPY, $65\%$ of $\hat\sigma$ comes from
$e_U$ (Section~\ref{sec:fitted}), so what the AL says about losses beyond VaR is mostly determined by the returns above
VaR. Under the AL itself this share is $95\%$, but there $e_L$ and $e_U$ are tied by
$e_U/e_L=(1-\tau)/\tau$; the SPY returns do not follow that ratio, and their share from
$e_L$, $35\%$, is about seven times $\tau$.

The paper makes three contributions. First, we show that whether the AL working
likelihood is adequate depends on the quantity being estimated: in the same fit, the
posterior for VaR is centred close to the historical VaR and the posterior for ES is far
from the historical CVaR, and we identify the two properties of the AL that cause the
difference and show that the error in ES is exactly the error in the AL's implied mean
(Corollary~\ref{cor:mean}).
Second, we define a diagnostic $\Rd$, the ratio of the VaR--CVaR gap implied
by a fitted AL to the true gap (Proposition~\ref{prop:R}), and give it in closed form for
common distributions (Section~\ref{sec:B}). Third, we document the failure in a simulation study, in a
cross-section of S\&P~500 stocks, and in a quantile regression with volatility
covariates.

Other ways of getting ES from an AL-based fit avoid \eqref{eq:es}. \citet{cgl2012}
estimate the AL's shape parameter, so that the location is no longer the $\tau$-quantile
(Section~\ref{sec:tail}). \citet{taylor2019} lets the scale vary over time and writes the
conditional ES as $\mu_t-\sigma_t/\tau$, with the conditional mean $\mu_t$ a small
constant estimated by the mean of the in-sample returns, so his ES does not use the AL's
mean. \citet{le2020} uses the same form with mixed-data-sampling dynamics and an AR(1)
conditional mean, and \citet{pzc2019}, who cite Taylor's article as Taylor (2017), write
that their theoretical results ``can be used to justify ex post the methods of his
paper''. \citet{sw2026} replace the AL by a mixture under the same constraint that the
$\tau$-quantile of the error is zero, which in their static model leaves four free
parameters where the AL with $\tau$ fixed has one.

Section~\ref{sec:setup} describes the model, the Gibbs sampler and the AL's expected
shortfall. Section~\ref{sec:mech} works out the properties of the AL shortfall and the
diagnostic $\Rd$. Section~\ref{sec:spy} has the results for SPY, Section~\ref{sec:A} the
simulation study, Section~\ref{sec:B} the values of $\Rd$ for other distributions and for
S\&P~500 stocks, Section~\ref{sec:C} the quantile regression, Section~\ref{sec:check}
the same programs run on data drawn from the AL and the residuals of the quantile
regression, and Section~\ref{sec:concl} concludes.

\section{Model and estimation}\label{sec:setup}

\subsection{Quantile regression and the AL working likelihood}\label{sec:ql}

We write $y_1,\dots,y_n$ for the daily returns and $\tau$ for the quantile level. Unless we
say otherwise, $\tau=0.05$. Let $Y$ be a return with a continuous distribution function $F$
and a finite mean. We work with returns rather than losses, so losses are in the lower
tail and, at the levels used in risk management, VaR and ES are negative numbers. The Value-at-Risk at level $\tau$ is the
$\tau$-quantile of the returns, and the expected shortfall is the mean return at or below
it,
\[
\VaR_\tau=F^{-1}(\tau)=\inf\{y:F(y)\ge\tau\},\qquad
\ES_\tau=\E[\,Y\mid Y\le\VaR_\tau\,].
\]
Because $F$ is continuous, $P(Y\le\VaR_\tau)=\tau$, and writing $Y=F^{-1}(U)$ with $U$
uniform on $(0,1)$ gives
\[
\ES_\tau=\frac{1}{\tau}\int_{-\infty}^{\VaR_\tau}y\,dF(y)=\frac{1}{\tau}\int_0^\tau F^{-1}(u)\,du,
\]
the average of the quantiles below level $\tau$. \citet{ru2000} call this quantity the
conditional Value-at-Risk (CVaR). They define it for a loss rather than a return, so for the
loss $-Y$ their CVaR at probability level $1-\tau$ is $-\ES_\tau$. We write ES for a value implied by a fitted
model and CVaR for the true value and its historical estimate, and call the distance
between VaR and either of them the VaR--CVaR gap.
The check loss is
$\rho_\tau(u)=u\,(\tau-\ind\{u<0\})$ and the AL density is
\begin{equation}
f(y\mid\omega,\sigma,\tau)=\frac{\tau(1-\tau)}{\sigma}
\exp\!\Big(-\frac{\rho_\tau(y-\omega)}{\sigma}\Big).
\label{eq:ald}
\end{equation}
This density puts mass $\tau$ below $\omega$, so $\omega$ is its $\tau$-quantile, and
maximising the likelihood in $\omega$ is the same as minimising
$\sum_i\rho_\tau(y_i-\omega)$. The quantile regression model is
$y_i=x_i^\top\beta+\varepsilon_i$ with $\varepsilon_i\sim\AL(0,\sigma,\tau)$. In
Sections~\ref{sec:spy}--\ref{sec:B} we only use an intercept, $x_i=1$, so $\omega=\beta$ is
a single VaR, and in Section~\ref{sec:C} we add covariates.

For the historical values, let $y_{(1)}\le\dots\le y_{(n)}$ be the sorted returns and
$k=\lfloor\tau n+\tfrac12\rfloor$, that is, $\tau n$ rounded to the nearest integer. The
historical VaR is $y_{(k)}$, the historical CVaR is $k^{-1}\sum_{i=1}^k y_{(i)}$, the mean
of the $k$ smallest returns, and $\tau_{\mathrm{eff}}=k/n$. For the SPY returns of
Section~\ref{sec:spy},\footnote{\label{fn:repo}The data, code and results files are at \repo. Every SPY number in Sections~\ref{sec:setup}--\ref{sec:spy} is computed by
\texttt{Spy\_run.py} and stored in \texttt{spy\_results.json}.} $\tau n=125.6$, so
$k=126$ and $\tau_{\mathrm{eff}}=0.050159$. Since $\tau n$ is not an integer,
$\sum_i\rho_\tau(y_i-\omega)$ has the unique minimiser $y_{(126)}$, so the
maximum-likelihood estimate of $\omega$ for SPY is the historical VaR.

\subsection{The Kozumi--Kobayashi Gibbs sampler}

The AL likelihood has no conjugate prior for $\beta$. \citet{kk2011} instead write the AL error as a
normal location--scale mixture,
\begin{equation}
\varepsilon_i=\theta v_i+\psi\sqrt{\sigma v_i}\,z_i,\qquad
v_i\sim\mathrm{Exp}(\sigma),\quad z_i\sim N(0,1),
\label{eq:kk}
\end{equation}
with $\theta=(1-2\tau)/(\tau(1-\tau))$ and $\psi^2=2/(\tau(1-\tau))$, so at $\tau=0.05$,
$\theta=18.947$ and $\psi^2=42.105$. Here $\mathrm{Exp}(\sigma)$ is the exponential
distribution with mean $\sigma$, that is, with density $\sigma^{-1}e^{-v/\sigma}$ for
$v>0$, and $N(m,V)$ is the normal distribution with mean $m$ and variance $V$ (a covariance
matrix when $m$ is a vector). Given
the latent $v_i$, $y_i$ is normal with mean $x_i^\top\beta+\theta v_i$ and variance
$\psi^2\sigma v_i$, and every full conditional distribution has a standard form.

We write $\mathrm{IG}(a,b)$ for the inverse gamma distribution with shape $a>0$ and scale
$b>0$, whose density is
\[
\frac{b^a}{\Gamma(a)}\,x^{-a-1}e^{-b/x},\qquad x>0,
\]
so that $1/X$ has the gamma distribution with shape $a$ and rate $b$, and $\E X=b/(a-1)$
for $a>1$. This is the parameterisation implied by the full conditional distribution of $\sigma$
in \citet{kk2011}. We write $\mathrm{IGauss}(m,\lambda)$ for the inverse Gaussian
distribution with mean $m$ and shape $\lambda$.

The priors are $\beta\sim N(\beta_0,B_0)$, with $\beta_0=0$ and $B_0$ the identity matrix,
and $\sigma\sim\mathrm{IG}(n_0/2,s_0/2)$ with $n_0=4$ and $s_0=0.024$, that is,
$\sigma\sim\mathrm{IG}(2,0.012)$, whose mean is $0.012$. Each iteration of the sampler
draws every $v_i$, then $\beta$, then $\sigma$ from their full conditional distributions,
\begin{align*}
1/v_i\mid\beta,\sigma,y&\sim\mathrm{IGauss}\big(d/c_i,\,d^2\big),\qquad
c_i^2=\frac{(y_i-x_i^\top\beta)^2}{\psi^2\sigma},\quad
d^2=\frac{2}{\sigma}+\frac{\theta^2}{\psi^2\sigma},\qquad i=1,\dots,n,\\
\beta\mid v,\sigma,y&\sim N(\tilde\beta,\tilde B),\qquad
\tilde B^{-1}=\sum_{i=1}^n\frac{x_ix_i^\top}{\psi^2\sigma v_i}+B_0^{-1},\\*
&\phantom{\sim N(\tilde\beta,\tilde B),\qquad}
\tilde\beta=\tilde B\Big(\sum_{i=1}^n\frac{x_i(y_i-\theta v_i)}{\psi^2\sigma v_i}
+B_0^{-1}\beta_0\Big),\\
\sigma\mid\beta,v,y&\sim\mathrm{IG}\Big(\frac{\tilde n}{2},\frac{\tilde s}{2}\Big),\qquad
\tilde n=n_0+3n,\quad
\tilde s=s_0+2\sum_{i=1}^n v_i+\sum_{i=1}^n\frac{(y_i-x_i^\top\beta-\theta v_i)^2}{\psi^2v_i}.
\end{align*}
These are the full conditional distributions of \citet{kk2011}, written in our notation
(their $p$ and $\tau$ are our $\tau$ and $\psi$; they state the first as a generalised
inverse Gaussian distribution for $v_i$, which is the same distribution). We run $4$ chains of
$15{,}000$ iterations with a burn-in of $5000$, which leaves $40{,}000$ draws of
$(\omega,\sigma)$, and we do not report the latent $v_i$. We check convergence with the potential scale reduction factor of \citet{gr1992} in
the form $\hat R=\sqrt{\hat\sigma^2/W}$, where $W$ is the average within-chain variance and
$\hat\sigma^2$ is from their eq.~(3), computed on split chains.

\subsection{Expected shortfall under the AL}\label{sec:es}

For $Y$ with a continuous
distribution and a finite mean, the expected check loss $\E\rho_\tau(Y-\xi)$ is minimised
at $\xi=\VaR_\tau$, and its minimum is $\tau(\E Y-\ES_\tau)$
\citep[Theorem~2 and its proof]{bkk2004}. \citet[eqs.~(7)--(9)]{mpr2021} write this as
\begin{equation}
\ES_\tau=\E Y-\frac{\delta}{\tau},\qquad
\delta=\E\big[(Y-\VaR_\tau)\big(\tau-\ind\{Y<\VaR_\tau\}\big)\big]=\E\rho_\tau(Y-\VaR_\tau),
\label{eq:delta}
\end{equation}
and, following \citet{taylor2019}, note that the scale parameter of an AL equals $\delta$.

\begin{lemma}\label{lem:delta}
If $Y\sim\AL(\omega,\sigma,\tau)$, then $\VaR_\tau=\omega$ and
$\delta=\E\rho_\tau(Y-\omega)=\sigma$.
\end{lemma}

\begin{proof}
The density \eqref{eq:ald} puts mass $\tau$ below $\omega$, so $\VaR_\tau=\omega$. Since
\eqref{eq:ald} integrates to one for every $\sigma>0$,
\[
\int_{-\infty}^{\infty}\exp\Big(-\frac{\rho_\tau(y-\omega)}{\sigma}\Big)\,dy=\frac{\sigma}{\tau(1-\tau)}.
\]
Differentiating both sides in $\sigma$ gives
$\int\rho_\tau(y-\omega)\,\sigma^{-2}\exp\{-\rho_\tau(y-\omega)/\sigma\}\,dy=1/(\tau(1-\tau))$,
and multiplying both sides by $\tau(1-\tau)\sigma$ gives $\E\rho_\tau(Y-\omega)=\sigma$.
\end{proof}

The mean of the AL is $\mu_{\AL}=\omega+\theta\sigma$, from taking expectations in
\eqref{eq:kk}: $\E\varepsilon=\theta\,\E v=\theta\sigma$. Putting $\delta=\sigma$ and
$\E Y=\mu_{\AL}$ in \eqref{eq:delta} gives
\[
\ES_\tau=\mu_{\AL}-\frac{\sigma}{\tau}
=\omega+\sigma\Big(\frac{1-2\tau}{\tau(1-\tau)}-\frac{1}{\tau}\Big)
=\omega-\frac{\sigma}{1-\tau},
\]
since $(1-2\tau)-(1-\tau)=-\tau$. This is \eqref{eq:es}, and each posterior draw gives
$\ES^{(g)}=\omega^{(g)}-\sigma^{(g)}/(1-\tau)$. The same formula follows from the lower
tail: below $\omega$ the AL density \eqref{eq:ald} is proportional to
$e^{-(1-\tau)(\omega-y)/\sigma}$, so given $Y<\omega$ the shortfall $\omega-Y$ is
exponential with mean $\sigma/(1-\tau)$. We come back to the form $\mu_{\AL}-\sigma/\tau$ in
Corollary~\ref{cor:mean}.

\subsection{An exact posterior for the intercept-only model}

When $x_i=1$, the likelihood is $\{\tau(1-\tau)\}^n\sigma^{-n}\exp\{-S(\omega)/\sigma\}$,
where $S(\omega)=\sum_i\rho_\tau(y_i-\omega)$. With the $\mathrm{IG}(n_0/2,s_0/2)$ prior
this gives $\sigma\mid\omega,y\sim\mathrm{IG}\big(n+n_0/2,\;S(\omega)+s_0/2\big)$, and
integrating $\sigma$ out leaves the one-dimensional posterior
\begin{equation}
\log p(\omega\mid y)=-\frac{\omega^2}{2B_0}
-\Big(n+\frac{n_0}{2}\Big)\log\!\Big(S(\omega)+\frac{s_0}{2}\Big)+\text{const},
\label{eq:exact}
\end{equation}
where $B_0=1$ is the prior variance of $\omega$. We sample this posterior independently by
inverse CDF on a grid, and then draw $\sigma$ from its inverse gamma distribution given
$\omega$.
This gives a check on the Gibbs sampler that does not involve a Markov chain. For SPY,
the two methods agree to within $1.28\times10^{-5}$ on the posterior means and 95\%
interval endpoints of $\omega$ and ES and on the posterior mean of $\sigma$. Both sets of
values are in the results file \texttt{spy\_results.json} in the GitHub repository of
footnote~\ref{fn:repo}, the Gibbs values under \texttt{posterior}, the exact values under
\texttt{exact} and the largest difference under \texttt{exact.worst\_gibbs\_diff}, and
\texttt{Spy\_run.py} prints them side by side in part~4 of its output.

\section{Properties of the AL shortfall}\label{sec:mech}

\subsection{The lower tail of the AL}

\begin{proposition}\label{prop:flat}
Under $\AL(\omega,\sigma,\tau)$, for every level $\tau'\le\tau$,
\[
\VaR_{\tau'}-\ES_{\tau'}=\frac{\sigma}{1-\tau}.
\]
\end{proposition}

\begin{proof}
Conditional on $Y<\omega$, $X=\omega-Y$ is exponential with rate
$\lambda=(1-\tau)/\sigma$. The exponential distribution is memoryless,
$P(X>a+t\mid X>a)=e^{-\lambda t}$, so the mean excess of $X$ over any threshold
$a\ge0$ is $1/\lambda$. For $\tau'\le\tau$, $\VaR_{\tau'}$ corresponds to such a
threshold.
\end{proof}

So under the AL, the expected loss beyond $\VaR_{\tau'}$ does not depend on $\tau'$, and
its ES is its VaR minus a constant that says nothing about the shape of the tail.

\subsection{The fitted scale}\label{sec:fitted}

By Proposition~\ref{prop:flat}, the AL's ES depends on the data only through $\omega$
and $\sigma$. For fixed $\hat\omega$, the maximum-likelihood estimate of $\sigma$ is the
mean check loss $\hat\sigma=S(\hat\omega)/n$ \citep{taylor2019}. The next
result splits it at $\hat\omega$.

\begin{lemma}\label{lem:sigma}
Let $\hat\omega=y_{(k)}$ be the $k$-th smallest observation, with no other observation
equal to it, and let $\tau=\tau_{\mathrm{eff}}=k/n$. Write $\VaR=y_{(k)}$ and $\CVaR$ for the
historical values of Section~\ref{sec:ql}, $\bar y_{\mathrm{up}}$ for the mean of the
$n-k$ observations above $\hat\omega$, $e_L=\VaR-\CVaR$ and $e_U=\bar y_{\mathrm{up}}-\VaR$.
Then
\begin{equation}
\hat\sigma=\tau(1-\tau)\,(e_L+e_U).
\label{eq:sigma}
\end{equation}
\end{lemma}

\begin{proof}
For $y_i\le\hat\omega$, $\rho_\tau(y_i-\hat\omega)=(1-\tau)(\hat\omega-y_i)$, and for
$y_i>\hat\omega$, $\rho_\tau(y_i-\hat\omega)=\tau(y_i-\hat\omega)$. The observations at
or below $\hat\omega$ are the $k$ smallest, so
\[
S(\hat\omega)=(1-\tau)\sum_{i=1}^{k}\big(\hat\omega-y_{(i)}\big)
+\tau\sum_{i=k+1}^{n}\big(y_{(i)}-\hat\omega\big).
\]
Since $\CVaR$ is the mean of the $k$ smallest observations, the first sum is
$k\hat\omega-\sum_{i\le k}y_{(i)}=k(\VaR-\CVaR)=k\,e_L$. Since $\bar y_{\mathrm{up}}$ is the
mean of the other $n-k$, the second sum is
$\sum_{i>k}y_{(i)}-(n-k)\hat\omega=(n-k)(\bar y_{\mathrm{up}}-\VaR)=(n-k)\,e_U$. Dividing
by $n$,
\[
\hat\sigma=\frac{S(\hat\omega)}{n}=(1-\tau)\,\frac{k}{n}\,e_L
+\tau\Big(1-\frac{k}{n}\Big)e_U,
\]
and putting $k/n=\tau$ gives $\hat\sigma=\tau(1-\tau)\,e_L+\tau(1-\tau)\,e_U$, which is
\eqref{eq:sigma}.
\end{proof}

At the nominal $\tau$ instead of $\tau_{\mathrm{eff}}$, \eqref{eq:sigma} holds only up to a
term of order $1/n$, which is why we report sample values at $\tau_{\mathrm{eff}}$. For SPY,
$e_L=0.010650$ is the mean distance below VaR of the $126$ returns at or below it, which
is the quantity $\VaR-\ES$ is supposed to estimate, and $e_U=0.019538$ is the mean excess
of the other $2386$ returns above VaR. So $35\%$ of $\hat\sigma$ comes from the lower tail
and $65\%$ from the returns above VaR, even though all of $\hat\sigma$ goes into the AL's
expected loss beyond VaR. Under the AL itself, $e_L=\sigma/(1-\tau)$ and
$e_U=\sigma/\tau$, so exactly $\tau$, here $5\%$, of $\hat\sigma$ comes from below VaR and
$95\%$ from above; for the AL data of Section~\ref{sec:aldata} the share from below is
$4.6\%$ (\texttt{al\_check/spy\_results.json} on GitHub, under
\texttt{sigma\_split.share\_lower}). So what goes wrong for SPY is not the large share from
above. The AL fixes $e_U/e_L$ at $(1-\tau)/\tau=19$, and the SPY returns do not follow that
ratio: their share from below, $35\%$, is about seven times $\tau$. The overshoot implied by the fitted AL, meaning its expected
loss beyond VaR, is $\hat\sigma/(1-\tau)=\tau(e_L+e_U)=0.001514$, while the realised value
is $e_L=0.010650$. These SPY values are in \texttt{spy\_results.json} on GitHub, under
\texttt{sigma\_split}.

With $\tau$ fixed, $\sigma$ is the AL's only scale: by \eqref{eq:ald} it sets both the rate
$(1-\tau)/\sigma$ of the lower tail and the rate $\tau/\sigma$ above $\omega$. So the lower
tail cannot be fitted on its own within the AL. That needs a parameter the AL with $\tau$
fixed does not have, such as a free shape parameter, which makes $\omega$ no longer the
$\tau$-quantile (Section~\ref{sec:tail}).

\subsection{A diagnostic ratio}

By the best-fitting AL we mean the one that minimises the expected negative
log-likelihood of the returns. Taking logs in \eqref{eq:ald},
$-\log f(Y\mid\omega,\sigma,\tau)=-\log\{\tau(1-\tau)\}+\log\sigma+\rho_\tau(Y-\omega)/\sigma$,
so, apart from the constant $-\log\{\tau(1-\tau)\}$, the expected negative log-likelihood is
\[
L(\omega,\sigma)=\log\sigma+\frac{\E\rho_\tau(Y-\omega)}{\sigma}.
\]
For every $\sigma>0$, $L$ is smallest at the $\omega$ that minimises $\E\rho_\tau(Y-\omega)$,
which is $\omega=\VaR_\tau$ (Section~\ref{sec:es}). Write
$\sigma^*=\E\rho_\tau(Y-\VaR_\tau)$. Then $L(\VaR_\tau,\sigma)=\log\sigma+\sigma^*/\sigma$
has derivative $(\sigma-\sigma^*)/\sigma^2$ in $\sigma$, which is negative below $\sigma^*$ and
positive above it, so $L$ is smallest at $\sigma=\sigma^*$. So $\sigma^*$ is the minimised
expected check loss, the population version of $\hat\sigma$ (Section~\ref{sec:fitted}).
Proposition~\ref{prop:R} below shows that the ratio of the overshoot $\sigma^*/(1-\tau)$
of the best-fitting AL to the true overshoot $e_L$ depends only on the mean, $\VaR_\tau$
and $\CVaR_\tau$. So the ratio can be computed for any continuous distribution with a
finite mean, without fitting a model.

\begin{proposition}\label{prop:R}
Let $Y$ have a continuous distribution with finite mean $\mu$, and let $e_L(\tau)=\VaR_\tau-\CVaR_\tau$ and
$e_U(\tau)=\E[\,Y\mid Y>\VaR_\tau]-\VaR_\tau$. The minimised expected check loss is
$\sigma^*=\tau(\mu-\CVaR_\tau)$ \citep[Theorem~2]{bkk2004}, and
\begin{equation}
\Rd(\tau)=\frac{\sigma^*/(1-\tau)}{e_L(\tau)}
=\frac{\tau(\mu-\CVaR_\tau)}{(1-\tau)(\VaR_\tau-\CVaR_\tau)}
=\tau\Big(1+\frac{e_U}{e_L}\Big).
\label{eq:R}
\end{equation}
\end{proposition}

\begin{proof}
By the law of total expectation over the events $\{Y\le\VaR_\tau\}$ and
$\{Y>\VaR_\tau\}$, which have probabilities $\tau$ and $1-\tau$ because $Y$ is continuous,
\[
\mu=\tau\,\E[\,Y\mid Y\le\VaR_\tau]+(1-\tau)\,\E[\,Y\mid Y>\VaR_\tau]
=\tau\,\CVaR_\tau+(1-\tau)(e_U+\VaR_\tau),
\]
by the definitions of $\CVaR_\tau$ and $e_U$. Subtracting $\CVaR_\tau$ gives
$\mu-\CVaR_\tau=(1-\tau)(e_L+e_U)$. Substituting this and $\sigma^*=\tau(\mu-\CVaR_\tau)$
into $\Rd(\tau)=\sigma^*/((1-\tau)e_L)$ gives the two expressions on the right.
\end{proof}

By Proposition~\ref{prop:flat}, $\sigma^*/(1-\tau)$ is the VaR--CVaR gap of the
best-fitting AL, so $\Rd(\tau)$ is the ratio of that gap to the true gap $e_L(\tau)$. It
does not change when $Y$ is replaced by $a+bY$ with $b>0$: then $\mu$, $\VaR_\tau$ and
$\CVaR_\tau$ become $a$ plus $b$ times their old values, so $\mu-\CVaR_\tau$ and
$\VaR_\tau-\CVaR_\tau$ are both multiplied by $b$ and the ratio in \eqref{eq:R} stays the
same. This is why a single closed-form value in Section~\ref{sec:B} covers a whole family,
such as the Normal, whatever its mean and standard deviation.

\begin{corollary}\label{cor:mean}
The expected shortfall \eqref{eq:es} of the best-fitting AL satisfies
$\ES_\tau-\CVaR_\tau=\mu_{\AL}-\mu$, where $\mu_{\AL}=\VaR_\tau+\theta\sigma^*$ is the
mean of that AL.
\end{corollary}

\begin{proof}
By Section~\ref{sec:es}, the ES of the best-fitting AL is $\mu_{\AL}-\sigma^*/\tau$, and
\eqref{eq:delta} applied to $Y$, with $\delta=\sigma^*$, gives
$\CVaR_\tau=\mu-\sigma^*/\tau$. Subtracting gives the result.
\end{proof}

So the AL gets the scale right, since its $\sigma$ is the $\delta$ of the returns, and its
ES is wrong by exactly the error in its mean, which the quantile and the scale fix at
$\VaR_\tau+\theta\sigma^*$, with $\theta=18.947$ at $\tau=0.05$. For SPY the posterior mean
of $\mu_{\AL}$ is $0.009926$, against a sample mean of $0.000580$, and the difference,
$0.0093$, is close to the distance of $0.0092$ between the posterior mean of ES and the
historical CVaR in Table~\ref{tab:spy}. These four values are in \texttt{spy\_results.json}
on GitHub, under \texttt{evidence.al\_implied\_mean}, \texttt{empirical.mean},
\texttt{evidence.mean\_gap} and \texttt{evidence.gap\_ES\_minus\_CVaR}. Using the sample mean in place of $\mu_{\AL}$
gives the estimator $\bar y-\hat\sigma/\tau$ of \citet{taylor2019}, which at
$\tau_{\mathrm{eff}}$ equals the historical CVaR exactly: then
$\bar y=\tau\,\CVaR+(1-\tau)\,\bar y_{\mathrm{up}}$, so
$\bar y-\CVaR=(1-\tau)(e_L+e_U)=\hat\sigma/\tau$ by Lemma~\ref{lem:sigma}.

\citet[Section~6]{srg2013} link $\sigma^*$ to the Bayesian fit. They outline an argument
that, when $\sigma$ has a prior supported on a compact interval, the posterior of
$(\omega,\sigma)$ concentrates at the true quantile and at the long-run average of the
expected check loss at that quantile, which for independent and identically distributed
returns is $\sigma^*$. Under their conditions, then, the posterior ES \eqref{eq:es}
concentrates at $\VaR_\tau-\sigma^*/(1-\tau)$, and $\Rd(\tau)$ is the large-sample ratio
of the VaR--CVaR gap implied by the posterior to the true gap. For SPY, the
posterior mean of $\sigma$ is $0.0014397$, compared with the maximum-likelihood value
$\hat\sigma=0.0014354$ at $\tau=0.05$, both in \texttt{spy\_results.json} on GitHub, under
\texttt{posterior.sigma\_mean} and \texttt{evidence.al\_mle.sigma\_hat\_nominal}.

If the returns follow $\AL(\omega,\sigma,\tau)$, then $\Rd(\tau)=1$ at that $\tau$,
because then $\sigma^*=\sigma$ by Lemma~\ref{lem:delta} and $e_L=\sigma/(1-\tau)$.

\section{Data and results for SPY}\label{sec:spy}

The data are daily simple returns on SPY, computed from adjusted closing prices, from
18 August 2015 to 13 August 2025, so $n=2512$.\footnote{The prices are from Yahoo Finance, downloaded with
\texttt{yfinance} by \texttt{Spy\_run.py}, and the returns are in
\texttt{spy\_returns\_2015\_2025.csv} at \repo.}
Table~\ref{tab:spy} shows the intercept-only fit at $\tau=0.05$ next to the historical
values. Its values are in \texttt{spy\_results.json} on GitHub, under \texttt{empirical},
\texttt{posterior} and \texttt{evidence}.

\begin{table}[!htb]
\centering\small
\caption{SPY, $n=2512$, $\tau=0.05$. Distances are in posterior standard
deviations of the quantity in that column.}
\label{tab:spy}
\begin{tabular}{lcc}
\toprule
 & VaR & ES / CVaR\\
\midrule
Historical ($k=126$) & $-0.017444$ & $-0.028094$\\
\midrule
AL posterior mean & $-0.017353$ & $-0.018868$\\
\quad 95\% credible interval & $[-0.017898,\ -0.016722]$ & $[-0.019417,\ -0.018235]$\\
Distance from historical & $0.29$ & $29.7$\\
Posterior draws reaching the historical value & -- & $0$ of $40{,}000$\\
\bottomrule
\end{tabular}
\end{table}

The historical VaR lies inside the 95\% credible interval for VaR, while the historical
CVaR lies far outside the interval for ES: it is $29.7$ posterior standard deviations from
the posterior mean, and none of the $40{,}000$ draws reaches it. In terms of the overshoot
beyond VaR, the AL implies $0.001515$, the posterior mean of $\sigma/(1-\tau)$, while the
realised mean overshoot is $0.010650$, $7.03$ times as large. These three values are in
\texttt{spy\_results.json} on GitHub, under \texttt{evidence.al\_mean\_excess},
\texttt{evidence.emp\_mean\_excess} and \texttt{evidence.excess\_ratio}.

Several checks indicate that the discrepancy does not come from the sampler, and
Section~\ref{sec:A} shows that it does not depend on the prior either. The split-chain
$\hat R$ is $1.00052$ for $\omega$ and $0.99999$ for $\sigma$, with effective sample sizes
of $2389$ and $19{,}484$ out of $40{,}000$, and the exact posterior \eqref{eq:exact} agrees
with the Gibbs output to within $1.28\times10^{-5}$. These values are in
\texttt{spy\_results.json} on GitHub, under \texttt{diagnostics} and
\texttt{exact.worst\_gibbs\_diff}.

\section{Simulation study}\label{sec:A}

To separate the effect of the likelihood from the effects of the sampler, the prior, the
sample size and the algebra of \eqref{eq:es}, we keep the estimation procedure fixed and
change only the data-generating process. There are five designs: an AL with the
maximum-likelihood parameters for SPY, under which the working likelihood is correct;
a Normal, a Student-$t(5)$ and a Student-$t(3)$ distribution with the sample mean $\bar y$
and sample standard deviation $s$ of SPY, that is, $N(\bar y,s^2)$ and $\bar y+s\sqrt{(\nu-2)/\nu}\,T_\nu$
with $T_\nu$ a standard Student-$t$ variable with $\nu$ degrees of freedom; and resampling with replacement from the SPY returns. Each design has
$1000$ replicates of size $n=2512$. The study uses the exact posterior
\eqref{eq:exact}, whose draws are independent, so there is no Markov chain and no
Markov chain error, and this many fits are feasible. On
one replicate from each design, the two methods agree to within $1.65\times10^{-5}$ on
the posterior mean and 95\% interval endpoints of ES; this largest difference is in
\texttt{experiment\_A\_results.json} on GitHub, under \texttt{A1.worst\_diff}.\footnote{The
code is \texttt{experiment\_a.py} and the results are in \texttt{experiment\_A\_results.json}
at \repo: Tables~\ref{tab:A2}, \ref{tab:A3} and \ref{tab:A4} under \texttt{A2}, \texttt{A3}
and \texttt{A4}.}

\begin{table}[!htb]
\centering\small
\caption{Coverage of 95\% credible intervals, $n=2512$, $1000$ replicates per
design. Miscentring is the average of (posterior mean ES $-$ true CVaR) in posterior
standard deviations.}
\label{tab:A2}
\begin{tabular}{lccc}
\toprule
design & ES coverage & VaR coverage & ES miscentring (sd)\\
\midrule
AL (working likelihood correct) & $95.9\%$ & $95.5\%$ & $0.0$\\
Normal & $0.0\%$ & $64.1\%$ & $15.7$\\
Student-$t(5)$ & $0.0\%$ & $60.0\%$ & $24.8$\\
Student-$t(3)$ & $0.0\%$ & $58.5\%$ & $32.5$\\
Bootstrap from SPY & $0.0\%$ & $61.2\%$ & $31.4$\\
\bottomrule
\end{tabular}
\end{table}

When the AL is the true model, the ES intervals reach their nominal coverage
(Table~\ref{tab:A2}). Under each of the other four designs their coverage is $0.0\%$,
because they are centred in the wrong place: on average the posterior mean of ES is $15.7$
to $32.5$ posterior standard deviations from the true CVaR. The VaR intervals also
under-cover in these designs, at $58.5\%$ to $64.1\%$, but because they are too narrow,
not because they are off-centre. On average their posterior mean is at most $0.42$
posterior standard deviations from the true VaR, while the posterior standard deviation is
only $0.42$ to $0.48$ times the standard deviation of the posterior mean across replicates.
The two quantities for each design are in \texttt{experiment\_A\_results.json} on GitHub,
under \texttt{var\_bias\_sd\_mean} and \texttt{var\_sd\_ratio} in \texttt{A2}.
This agrees with \citet{ywh2016}, who show, with $\sigma$ fixed, that under
misspecification the AL posterior for the quantile is centred on the usual quantile
regression estimator but its variance does not match the sampling variance of that
estimator.

ES coverage under misspecification does not get better with a larger sample
(Table~\ref{tab:A3}). In the four misspecified designs it is at most $9.8\%$ at $n=250$ and
$0.0\%$ at $n=1000$ and $n=2512$, while in the AL design it stays close to $95\%$.

\begin{table}[!htb]
\centering\small
\caption{ES coverage of 95\% credible intervals and ES miscentring (posterior mean ES
$-$ true CVaR, in posterior standard deviations) by sample size, $500$ replicates per
cell, separate from those of Table~\ref{tab:A2}. Growth is the miscentring at $n=2512$
divided by that at $n=250$; $\sqrt{2512/250}=3.17$.}
\label{tab:A3}
\begin{tabular}{lccccccc}
\toprule
 & \multicolumn{3}{c}{ES coverage} & \multicolumn{3}{c}{ES miscentring (sd)} & \\
\cmidrule(lr){2-4}\cmidrule(lr){5-7}
design & $n=250$ & $n=1000$ & $n=2512$ & $n=250$ & $n=1000$ & $n=2512$ & growth\\
\midrule
AL & $94.2\%$ & $95.6\%$ & $94.4\%$ & $0.0$ & $0.0$ & $0.0$ & --\\
Normal & $9.8\%$ & $0.0\%$ & $0.0\%$ & $5.3$ & $10.4$ & $15.7$ & $2.95$\\
Student-$t(5)$ & $1.6\%$ & $0.0\%$ & $0.0\%$ & $8.0$ & $15.4$ & $24.2$ & $3.01$\\
Student-$t(3)$ & $1.0\%$ & $0.0\%$ & $0.0\%$ & $10.9$ & $20.9$ & $32.2$ & $2.97$\\
Bootstrap & $1.0\%$ & $0.0\%$ & $0.0\%$ & $9.6$ & $19.5$ & $31.2$ & $3.25$\\
\bottomrule
\end{tabular}
\end{table}

The bias of \eqref{eq:es} does not go away as $n$ grows, while the
posterior standard deviation shrinks at rate $n^{-1/2}$, so the miscentring measured in
posterior standard deviations grows roughly like $\sqrt n$. From $n=250$ to $n=2512$ it
increases by a factor of $2.95$ to $3.25$ across the four misspecified designs, compared
with $\sqrt{2512/250}=3.17$ (last column of Table~\ref{tab:A3}).

We also tried $7$ priors, including $\beta\sim N(0,100)$ and ten-fold changes in $s_0$ in
either direction (Table~\ref{tab:A4}). Across them, the posterior mean of ES for SPY moves by
at most $0.14$ posterior standard deviations, while its distance from the historical CVaR
under the exact posterior is $29.9$ posterior standard deviations. Both values are in
\texttt{experiment\_A\_results.json} on GitHub, under \texttt{A4.prior\_max\_shift\_post\_sd}
and \texttt{A4.miscentring\_post\_sd}.

\begin{table}[!htb]
\centering\small
\caption{Prior sensitivity for SPY, exact posterior \eqref{eq:exact}. The prior is
$\beta\sim N(0,B_0)$ and $\sigma\sim\mathrm{IG}(2,s_0/2)$; the first row is ours. Shift is
the change in the posterior mean of ES from the first row, in posterior standard deviations
of ES ($0.000309$).}
\label{tab:A4}
\begin{tabular}{lccc}
\toprule
$s_0$ & $B_0$ & posterior mean of ES & shift (sd)\\
\midrule
$0.024$ & $1$ & $-0.018863$ & $0.000$\\
$0.024$ & $100$ & $-0.018863$ & $0.000$\\
$0.024$ & $0.01$ & $-0.018863$ & $0.001$\\
$0.012$ & $1$ & $-0.018861$ & $0.008$\\
$0.048$ & $1$ & $-0.018868$ & $-0.016$\\
$0.0024$ & $1$ & $-0.018859$ & $0.014$\\
$0.24$ & $1$ & $-0.018908$ & $-0.144$\\
\bottomrule
\end{tabular}
\end{table}

\section{The diagnostic across distributions and assets}\label{sec:B}

In this section we evaluate the diagnostic $\Rd(\tau)$ from Proposition~\ref{prop:R}, the
ratio of the VaR--CVaR gap implied by a fitted AL, $\sigma^*/(1-\tau)$, to the true gap
$\VaR_\tau-\CVaR_\tau$. A value of $1$ means the AL's ES at level $\tau$ is correct, and a
value below $1$ is the fraction of the true gap that the AL's gap represents.

\subsection{Closed-form values}\label{sec:cf}

Table~\ref{tab:B0} gives $\Rd(\tau)$ for SPY and for three reference
distributions.\footnote{The code for Sections~\ref{sec:cf} and \ref{sec:tail} is
\texttt{experiment\_b\_overall.py} and the results are in \texttt{experiment\_B\_results.json}
at \repo: Table~\ref{tab:B0} under \texttt{B0}.} For
SPY the AL represents $0.1422$ of the true gap at $\tau=0.05$ and $0.0324$ of
it at $\tau=0.01$. In every row, $\Rd$ decreases as $\tau$ decreases. So the failure seen
at $\tau=0.05$ is not specific to that level, and it gets worse further into the tail: the
smaller $\tau$ is, the smaller the part of the true gap that the AL's ES represents, and the
more it understates the expected loss beyond VaR.

\begin{table}[!htb]
\centering\small
\caption{$\Rd(\tau)$. The SPY row is the sample value at $\tau_{\mathrm{eff}}$
($n=2512$); the other rows are closed forms. An AL whose parameter equals the
column's $\tau$ has $\Rd=1$.}
\label{tab:B0}
\begin{tabular}{lccc}
\toprule
$\tau$ & 0.01 & 0.05 & 0.10\\
\midrule
SPY & $0.0324$ & $0.1422$ & $0.2463$\\
Normal & $0.0794$ & $0.2598$ & $0.4119$\\
Student-$t(5)$ & $0.0414$ & $0.1738$ & $0.3096$\\
Student-$t(3)$ & $0.0287$ & $0.1341$ & $0.2540$\\
\bottomrule
\end{tabular}
\end{table}

\subsection{Tail weight}\label{sec:tail}

\citet{sw2026} explain the AL's poor fit to returns by the shape of the return
distribution, writing that the AL ``does not fit the distribution of asset returns,
which is characterized by sharp peaks and heavy tails.'' As a statement about density
fit this is correct, and $\Rd$ does move in the direction it suggests: in the Student-$t$
family at $\tau=0.05$, $\Rd$ is $0.1187$ for $\nu=2.5$, $0.1341$ for
$\nu=3$, $0.1738$ for $\nu=5$, $0.2127$ for $\nu=10$ and $0.2547$
for $\nu=100$ (in \texttt{experiment\_B\_results.json} on GitHub, under \texttt{B4}), and
for the Normal it is $0.2598$ (Table~\ref{tab:B0}). So for Gaussian
data, which have neither a sharp peak nor a heavy tail, the AL represents about a quarter
of the VaR--CVaR gap. Tail weight therefore changes the size of the discrepancy, but not
its cause, which is Proposition~\ref{prop:flat} and Lemma~\ref{lem:sigma}.

With its shape fixed at $\tau$, the AL has a single scale (Section~\ref{sec:fitted}), so
the discrepancy cannot be removed within that model. \citet{cgl2012} instead write the AL with a shape parameter $p$ in place of $\tau$ in
\eqref{eq:ald} and estimate it together with the location and scale. Fitting that AL to SPY
by maximum likelihood gives $\hat p=0.5051$, and the gap between its 5\% VaR and ES, both
computed from the fitted distribution, is $0.7061$ of the historical VaR--CVaR gap, against
$0.1422$ for the AL with $p=\tau$. These three values are in
\texttt{experiment\_B\_results.json} on GitHub, under \texttt{B7}. A free shape parameter therefore recovers a large part of
the gap, but the location is then the $\hat p$-quantile, not the $\tau$-quantile, so the fit
is no longer a quantile regression at level $\tau$.

\subsection{Cross-section of S\&P~500 stocks}\label{sec:B6}

The empirical results so far use SPY only. To check that the failure is not specific to
SPY, we compute $\Rd$ for many assets, which is cheap because $\Rd$ needs only the sample
mean, the $\tau$-quantile and the mean below it. We
downloaded adjusted closing prices for SPY and the S\&P~500 constituents as listed on
27 September 2026, over the same window as the SPY sample, and kept every series with at
least $750$ returns. That gives $492$ series: SPY and $491$ stocks. Of these, $30$ have
fewer returns than SPY, and the shortest has $894$; the number of returns of each series
is the column \texttt{n} of \texttt{B6\_mine\_by\_asset.csv} on GitHub.

Table~\ref{tab:B6} summarises $\Rd$ across the $492$ series at $\tau=0.01$, $0.05$ and
$0.10$.\footnote{The value of $\Rd$ for each of the $492$ series at each
level is in \texttt{B6\_mine\_by\_asset.csv} and the summary is in \texttt{B6\_mine.json}.
Both are written by \texttt{experiment\_b6.py} from the saved prices in
\texttt{B6\_mine\_panel.csv}, and the list of tickers is in
\texttt{sp500\_tickers\_20260927.txt}, at \repo.}

\begin{table}[!htb]
\centering\small
\caption{$\Rd$ across $492$ series (SPY and $491$ current S\&P~500 constituents). Median $k$
is the median number of returns at or below VaR. The last column is the closed-form value for
the Normal (Table~\ref{tab:B0}).}
\label{tab:B6}
\begin{tabular}{lccccc}
\toprule
$\tau$ & median $k$ & median $\Rd$ & largest $\Rd$ & series with $\Rd>1$ & Normal\\
\midrule
0.01 & $25$ & $0.0315$ & $0.0677$ & $0$ & $0.0794$\\
0.05 & $126$ & $0.1446$ & $0.2180$ & $0$ & $0.2598$\\
0.10 & $251$ & $0.2652$ & $0.3802$ & $0$ & $0.4119$\\
\bottomrule
\end{tabular}
\end{table}

At each level, even the largest value of $\Rd$ is below the Gaussian value, and so below
$1$: at $\tau=0.05$ the largest is $0.2180$, against $0.2598$ for the Normal. So for every
one of the $492$ series the AL represents a smaller part of the true gap than it would for
Gaussian returns, and the AL's ES understates the expected loss beyond VaR. At $\tau=0.05$
the median, $0.1446$, lies between the closed-form values for $t(3)$, $0.1341$, and the
Normal. SPY sits at the 55th, 42nd and 11th percentiles of the cross-section at
$\tau=0.01$, $0.05$ and $0.10$, which can be read off the columns \texttt{R\_0.01},
\texttt{R\_0.05} and \texttt{R\_0.1} of \texttt{B6\_mine\_by\_asset.csv} on GitHub.
The series are today's constituents, so companies that left the index during the sample
period are missing, and they share one market history, so they are not independent.

\section{A conditional quantile model}\label{sec:C}

So far we have used the intercept-only model, where the VaR is the same on every day. In
this section we fit the quantile regression
\begin{equation}
\VaR_t=b_0+b_1\,|r_{t-1}|+b_2\,\mathrm{sd}_{20}(r)_t,
\label{eq:C}
\end{equation}
where $\mathrm{sd}_{20}(r)_t$ is the standard deviation of the previous $20$
returns. Both covariates use data up to $t-1$ only and are standardised. Because of the
rolling window the sample has $n=2492$ returns, from 16 September 2015 to 13 August 2025,
instead of $2512$. The model is
estimated with the sampler and prior settings of Section~\ref{sec:setup}, with
$\beta=(b_0,b_1,b_2)$ and $\beta\sim N(0,I_3)$, that is, independent $N(0,1)$ priors on
the three coefficients.\footnote{The code is \texttt{experiment\_c.py} and the results are
in \texttt{experiment\_C\_results.json} at \repo: Table~\ref{tab:C2} under \texttt{C2}.}

Both slope coefficients are negative, and their 95\% credible intervals do not include
zero: $b_1=-0.003826$ $[-0.004521,\ -0.003049]$ and $b_2=-0.006754$ $[-0.007707,\ -0.005806]$
per standard deviation of the covariate. The largest $\hat R$ across parameters is
$1.0039$. The fitted $\VaR_t$ is exceeded on $4.94\%$ of days, against the nominal $5\%$,
and it ranges from $-0.10208$ to $-0.00541$, a factor of $18.9$. These values are in
\texttt{experiment\_C\_results.json} on GitHub, under \texttt{C1}.

Since $\sigma$ does not depend on $t$, by Proposition~\ref{prop:flat} the model gives
every one of the $2492$ days the same expected overshoot,
$\VaR_t-\ES_t=\sigma/(1-\tau)=0.0012900$, with 95\% credible interval
$[0.0012406,\ 0.0013412]$ (\texttt{C2.al\_overshoot} and \texttt{C2.al\_ci} in
\texttt{experiment\_C\_results.json} on GitHub). Table~\ref{tab:C2} compares this value with the realised mean
overshoot of the $123$ exceedances, grouped by the quartile of the volatility covariate
on the day of the exceedance (quartiles over all $2492$ days).

\begin{table}[!htb]
\centering\small
\caption{Realised mean overshoot beyond $\VaR_t$ by volatility quartile, from calm to
volatile, with 95\% bootstrap intervals. The AL asserts $0.0012900$
$[0.0012406,\ 0.0013412]$ in every quartile. $n=2492$.}
\label{tab:C2}
\begin{tabular}{lccc}
\toprule
quartile & exceedances & realised overshoot [95\%] & ratio to AL\\
\midrule
1 & $30$ & $0.00711$ $[0.00482,\ 0.00967]$ & $5.51$\\
2 & $25$ & $0.00620$ $[0.00418,\ 0.00845]$ & $4.81$\\
3 & $38$ & $0.00518$ $[0.00375,\ 0.00673]$ & $4.02$\\
4 & $30$ & $0.01161$ $[0.00734,\ 0.01636]$ & $9.00$\\
\bottomrule
\end{tabular}
\end{table}

In every quartile, the bootstrap interval for the realised overshoot lies above the
model's credible interval, with ratios from $4.02$ to $9.00$ (Table~\ref{tab:C2}).

\paragraph{Comparison with the intercept-only model.} Fitted to the same $2492$ returns,
the intercept-only model asserts an overshoot of $0.0014991$, and the realised mean
overshoot beyond its VaR is $0.010775$, a ratio of $7.19$. With the covariates the two
are $0.0012900$ and $0.007427$, a ratio of $5.76$. These values are in
\texttt{experiment\_C\_results.json} on GitHub, under \texttt{C5}. The asserted overshoot is smaller
because the intercept-only model is nested in \eqref{eq:C}, so the minimised check loss
of \eqref{eq:C}, which sets $\hat\sigma$ (Section~\ref{sec:fitted}), cannot be larger.
The realised overshoot is smaller because it is now measured from a VaR that moves with
volatility, and so is lower on volatile days. So the covariates bring the AL closer to the realised
overshoot, but it still understates it by a factor of about six, and it gives every day the
same overshoot.

\section{Checks with data drawn from the AL}\label{sec:check}

\subsection{The programs on AL data}\label{sec:aldata}

We have attributed the results of Sections~\ref{sec:spy}--\ref{sec:C} to the AL not being
the distribution of the returns. If that is right, and the theory and the programs have no
errors, then on data that really come from an AL the same programs should find no
problem. To check this we drew $2512$ returns from $\AL(\omega,\sigma,0.05)$ with
$\omega=-0.017444$ and $\sigma=0.0014383$, the AL fitted to SPY (the maximum-likelihood
values at $\tau_{\mathrm{eff}}$, which also define the AL design of Section~\ref{sec:A}).
For the $492$ series of Section~\ref{sec:B6} we drew, for each one, the same number of
returns from the AL fitted to that series. We then ran every program behind
Tables~\ref{tab:spy}--\ref{tab:C2} on these data, with the same settings as for the real
data.\footnote{The data are made by \texttt{make\_al\_data.py} and are in the folder
\texttt{al\_check} at \repo, with the AL parameters in \texttt{al\_data.json}. The programs
are run in that folder, so each writes its results file there under the same name as for
the real data.} Table~\ref{tab:check} shows the main result of each table for the real
data and for the AL data, next to the value expected when the AL is the true model.

\begin{table}[!htb]
\centering\small
\caption{The main results of Tables~\ref{tab:spy}--\ref{tab:C2} for the real data and for
data drawn from the AL. The real data are SPY, except in the row for Table~\ref{tab:B6},
which is over the $492$ series. The last column is the value expected if the AL is the
true model; a dash means there is no single expected value.}
\label{tab:check}
\begin{tabular}{@{}cp{6.4cm}ccc@{}}
\toprule
Table & result & real data & AL data & expected\\
\midrule
\ref{tab:spy} & distance of the posterior mean of ES from the historical CVaR, in posterior
standard deviations & $29.7$ & $0.76$ & --\\
\ref{tab:spy} & realised overshoot beyond VaR divided by the overshoot the AL implies
 & $7.03$ & $0.92$ & $1$\\
\ref{tab:A2} & coverage of the 95\% ES intervals, bootstrap design & $0.0\%$ & $89.7\%$
 & $95\%$\\
\ref{tab:B0} & $\Rd$ at $\tau=0.05$ & $0.1422$ & $1.0901$ & $1$\\
\ref{tab:B6} & median $\Rd$ at $\tau=0.05$ & $0.1446$ & $1.0023$ & $1$\\
\ref{tab:C2} & realised overshoot divided by the AL's, in the four volatility quartiles
 & $4.02$ to $9.00$ & $0.72$ to $1.16$ & $1$\\
\bottomrule
\end{tabular}
\end{table}

In every row the real data show the failure and the AL data do not. The first two rows
are from Table~\ref{tab:spy}. If the AL is the true model, the overshoot beyond VaR that it
implies, $\sigma/(1-\tau)$, is the true one (Proposition~\ref{prop:flat}), so the ratio in
the second row should be $1$. The distance in the first row has no single expected value:
even then, the posterior mean of ES and the historical CVaR are two different estimates
from the same returns, so they differ by a small random amount, and the distance should
only be small. For SPY the posterior mean of ES is $29.7$ posterior
standard deviations from the historical CVaR, and the realised overshoot beyond VaR is
$7.03$ times what the AL implies. For the AL data the distance is $0.76$ and the ratio is
$0.92$, close to $1$. So the same code gets ES right when the AL is the right model, and
the error for SPY comes from the SPY returns not following the AL.

The third row is from Table~\ref{tab:A2}. The bootstrap design makes new samples by
resampling one return series. If the AL is the true model, 95\% intervals should contain
the true value $95\%$ of the time. Resampling SPY gives ES intervals that never contain the
true value, while resampling the AL series gives intervals that contain it $89.7\%$ of the
time. This is not quite $95\%$ because the one AL sample being resampled differs a little,
by chance, from the AL it was drawn from: its $\Rd$ is $1.0901$ rather than $1$. So the
$0.0\%$ for SPY comes from the shape of the SPY returns, not from the bootstrap or the
code.

The next two rows are from Tables~\ref{tab:B0} and \ref{tab:B6}. $\Rd$ is the fraction of
the true VaR--CVaR gap that the AL's ES represents (Proposition~\ref{prop:R}), and it is
$1$ when the AL is the true model. It is $1.0901$ for the AL series and has median $1.0023$
over the $492$ AL series, while for SPY and the real series it is $0.1422$ and $0.1446$, so
there the AL's ES represents only about a seventh of the true gap.

The last row is from the quantile regression of Table~\ref{tab:C2}. If the AL model is
true, the overshoot beyond each day's VaR is $\sigma/(1-\tau)$ on every day, so the ratio
should be $1$ in each quartile. With the AL data the
realised overshoot is $0.72$ to $1.16$ times the AL's in the four volatility quartiles,
against $4.02$ to $9.00$ for SPY. So the failure in Section~\ref{sec:C} does not come from
the regression code either.

The values for the AL data are in the folder \texttt{al\_check} on GitHub, in results
files with the same names and keys as for the real data. The programs behind
Tables~\ref{tab:A3} and \ref{tab:A4} were run as well, and their results are in the same
folder. Together these runs support our explanation: the failure comes from the AL not
fitting the returns, not from an error in the theory or in the programs.

\subsection{Residuals of the quantile regression}\label{sec:resid}

\begin{figure}[!htb]
\centering
\includegraphics[width=\textwidth]{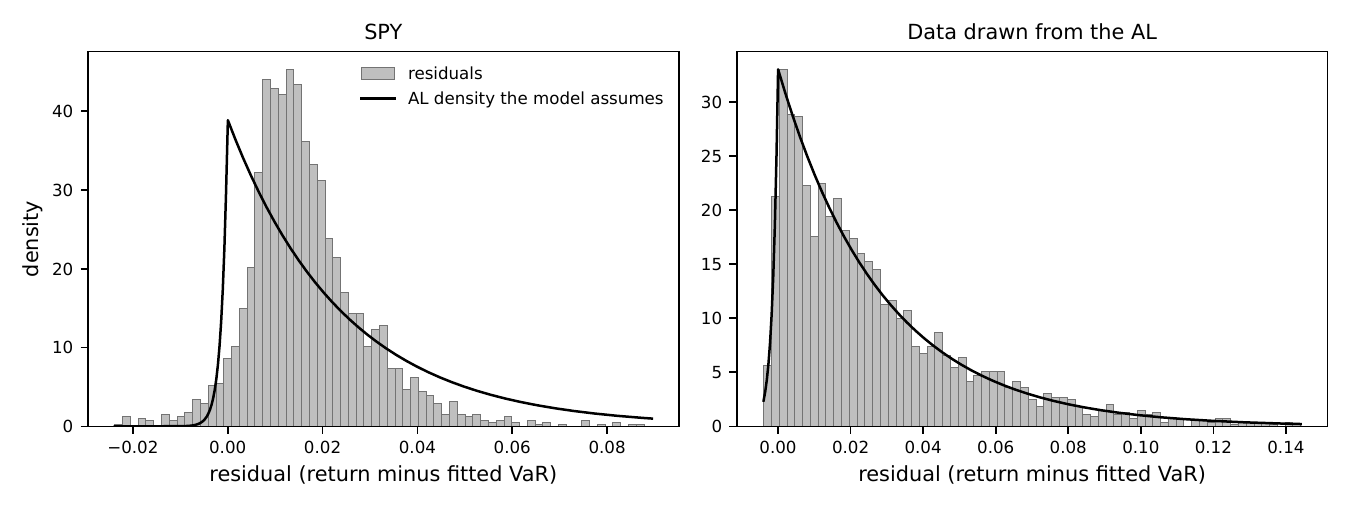}
\caption{Residuals of the quantile regression \eqref{eq:C} at $\tau=0.05$, that is, each
day's return minus its fitted VaR (bars), and the density of the AL that the model assumes
for them (curve). Left: SPY. Right: data drawn from the AL (Section~\ref{sec:aldata}). If
the AL is right, the bars follow the curve. The smallest $0.2\%$ and the largest $0.5\%$ of
the residuals are outside the plotted range.}
\label{fig:resid}
\end{figure}

Section~\ref{sec:aldata} checks the programs. A more direct way to see whether the AL is a
reasonable model for the errors is to look at the errors themselves. We fitted the
quantile regression \eqref{eq:C} at $\tau=0.05$ by linear programming, which gives the
same coefficients as maximising the AL likelihood, and computed the residuals
$e_i=y_i-x_i^\top\hat\beta$, that is, each day's return minus that day's fitted VaR. A
positive residual means the return was above VaR, and a negative one is a loss beyond VaR
by that amount. If the AL were the right model, the residuals would follow
$\AL(0,\hat\sigma,0.05)$, where $\hat\sigma$, the maximum-likelihood estimate of $\sigma$,
is their mean check loss. Figure~\ref{fig:resid} shows the histogram of the residuals
(bars) with this AL density (curve), for SPY and for the AL data of
Section~\ref{sec:aldata}. If the AL is right, the bars should follow the
curve.\footnote{The code is \texttt{residual\_hist.py} at \repo. It writes the values to
\texttt{residual\_results.json} and the figure to \texttt{residual\_hist.pdf}.}

For the AL data the bars follow the curve. For SPY they do not. The curve has a tall peak
at $0$, falls almost straight down on the left, and has a long, slowly falling tail on the
right. The SPY bars peak above $0$, their right side is shorter than the curve's, and their
left side, where the losses beyond VaR are, reaches much further than the curve. The
negative residuals average $-0.007277$, $5.7$ times the $-\hat\sigma/(1-\tau)=-0.001285$
that the AL gives. The true ES depends only on this left part, so the figure shows directly
why the AL's ES understates the loss beyond VaR. These values are in
\texttt{residual\_results.json} on GitHub, under \texttt{SPY}. So the AL is not a reasonable
distribution for the errors of the SPY returns, and it fails on the side that ES depends on.

\section{Conclusion}\label{sec:concl}

Under mild conditions, the posterior from an AL working likelihood concentrates on the
true quantile even when the AL is misspecified \citep[for fixed $\sigma$]{srg2013}, and
the AL's closed-form expected shortfall seems to extend the method to ES. We find
that the ES obtained this way is miscentred for SPY and in the misspecified
simulation designs, for two reasons that are properties of the AL itself. The AL's lower
tail is exponential, so its ES differs from its VaR by the constant $\sigma/(1-\tau)$, and
the fitted $\sigma$ is the minimised mean check loss, which is mostly determined by the
returns above VaR. For daily SPY returns the posterior for ES is $29.7$ posterior standard
deviations from the historical CVaR, while the posterior for VaR is consistent with the
historical VaR. In the simulation study the ES credible intervals had zero coverage under
every misspecified design at the SPY sample size, and the coverage did not get better with
a larger sample. The diagnostic $\Rd$, the fraction of the true VaR--CVaR gap that a
fitted AL can represent, is below one for all $492$ series in an S\&P~500 cross-section.
Adding volatility covariates to the quantile reduces the discrepancy, but the asserted
overshoot remains the same on every day, and the realised mean overshoot is still about
six times as large. Run on data drawn from the AL, the same programs do not find the
discrepancy, and the residuals of the quantile regression for SPY are far from the fitted
AL.

We leave several extensions for future work. Our Bayesian analysis uses a single
index and is in-sample, so an out-of-sample comparison over a rolling window would show
whether the discrepancy affects ES forecasts in the same way. Our models keep the AL
scale constant over time; \citet{taylor2019} lets it vary, and how a time-varying scale
behaves inside a Bayesian fit is a further question. Finally, since the discrepancy
comes from the AL with $\tau$ fixed having a single scale parameter for both sides of the
quantile (Section~\ref{sec:fitted}), a natural next step is Bayesian inference with a working
likelihood that keeps the quantile property but gives the lower tail its own parameter,
such as the mixture of \citet{sw2026}.

\section*{Acknowledgements}
Claude (Anthropic) drafted the code and drafted and edited the text. We optimised the code to run faster, removed parts that were not needed, checked that the outputs were correct and in the right format, and take full responsibility for the content of this paper.

\setlength{\bibsep}{2pt plus 0.3ex}


\begin{thebibliography}{99}\small

\bibitem[Basel Committee on Banking Supervision(2016)]{basel2016}
Basel Committee on Banking Supervision (2016). \emph{Minimum capital requirements for
market risk}. Bank for International Settlements, January 2016.

\bibitem[Bassett et~al.(2004)]{bkk2004}
Bassett, G.~W., Koenker, R. and Kordas, G. (2004). Pessimistic portfolio allocation
and Choquet expected utility. \emph{Journal of Financial Econometrics} 2(4), 477--492.

\bibitem[Chen et~al.(2012)]{cgl2012}
Chen, Q., Gerlach, R. and Lu, Z. (2012). Bayesian Value-at-Risk and expected shortfall
forecasting via the asymmetric Laplace distribution. \emph{Computational Statistics \&
Data Analysis} 56(11), 3498--3516.

\bibitem[Gelman and Rubin(1992)]{gr1992}
Gelman, A. and Rubin, D.~B. (1992). Inference from iterative simulation using multiple
sequences. \emph{Statistical Science} 7(4), 457--472.

\bibitem[Koenker and Bassett(1978)]{kb1978}
Koenker, R. and Bassett, G. (1978). Regression quantiles. \emph{Econometrica} 46(1),
33--50.

\bibitem[Kozumi and Kobayashi(2011)]{kk2011}
Kozumi, H. and Kobayashi, G. (2011). Gibbs sampling methods for Bayesian quantile
regression. \emph{Journal of Statistical Computation and Simulation} 81(11),
1565--1578.

\bibitem[Le(2019)]{le2020}
Le, T.~H. (2019). Forecasting Value at Risk and expected shortfall with mixed data
sampling. Working paper, version of 24 April 2019, SSRN 3509425,
\url{https://ssrn.com/abstract=3509425}.

\bibitem[Merlo et~al.(2021)]{mpr2021}
Merlo, L., Petrella, L. and Raponi, V. (2021). Forecasting VaR and ES using a joint
quantile regression and implications in portfolio allocation. arXiv:2106.06518.

\bibitem[Patton et~al.(2017)]{pzc2019}
Patton, A.~J., Ziegel, J.~F. and Chen, R. (2017). Dynamic semiparametric models for
expected shortfall (and Value-at-Risk). arXiv:1707.05108.

\bibitem[Rockafellar and Uryasev(2000)]{ru2000}
Rockafellar, R.~T. and Uryasev, S. (2000). Optimization of conditional value-at-risk.
\emph{Journal of Risk} 2(3), 21--41.

\bibitem[Shi and Wang(2026)]{sw2026}
Shi, S. and Wang, X. (2026). Parametric quantile regression using mixture
distributions: estimating Value at Risk and expected shortfall. \emph{Journal of
Forecasting} 45, 1495--1510.

\bibitem[Sriram et~al.(2013)]{srg2013}
Sriram, K., Ramamoorthi, R.~V. and Ghosh, P. (2013). Posterior consistency of Bayesian
quantile regression based on the misspecified asymmetric Laplace density.
\emph{Bayesian Analysis} 8(2), 479--504. Correction: \emph{Bayesian Analysis} 12(4)
(2017), 1217--1219.

\bibitem[Taylor(2019)]{taylor2019}
Taylor, J.~W. (2019). Forecasting Value at Risk and expected shortfall using a
semiparametric approach based on the asymmetric Laplace distribution. \emph{Journal of
Business \& Economic Statistics} 37(1), 121--133.

\bibitem[Yang et~al.(2016)]{ywh2016}
Yang, Y., Wang, H.~J. and He, X. (2016). Posterior inference in Bayesian quantile
regression with asymmetric Laplace likelihood. \emph{International Statistical Review}
84(3), 327--344, doi:10.1111/insr.12114.

\bibitem[Yu and Moyeed(2001)]{ym2001}
Yu, K. and Moyeed, R.~A. (2001). Bayesian quantile regression. \emph{Statistics \&
Probability Letters} 54(4), 437--447.

\end{thebibliography}
\end{document}